\documentclass[12pt]{iopart}

\usepackage{graphicx}

\begin{document}

\title{Saving Entanglement via Nonuniform Sequence of $\pi$ Pulses}
\author{G. S. Agarwal\footnote{Material presented in the lecture at the
International Conference ``Quantum Nonstationary Systems'', at
Brasilia, Brazil, Oct 2009.}}
\address{Department of Physics, Oklahoma State University,
Stillwater, OK - 74078, USA} \eads{\mailto{agirish@okstate.edu}}
\date{\today}

\begin{abstract}
We examine the question of survival of quantum entanglement between
the bipartite states and multiparticle states like GHZ states under
the action of a dephasing bath by the application of sequence of
$\pi$ pulses. We show the great advantage of the pulse sequence of
Uhrig [ 2007 {\it Phys. Rev. Lett.} {\bf 98} 100504] applied at
irregular intervals of time, in controlling quantum entanglement. In
particular death of entanglement could be considerably delayed by
pulses. We use quantum optical techniques to obtain exact results.
\end{abstract}

\vspace*{1in} \noindent{Contents}
\begin{enumerate}
\item[1.] Introduction\hspace{\fill}2
\item[2.] Dynamical decay of entanglement under dephasing\hspace{\fill}3
\item[3.] Single qubit coherence $S(t)$\hspace{\fill}5
\item[4.] Saving entanglement: numerical results\hspace{\fill}6
\item[5.] Conclusions\hspace{\fill}8
\end{enumerate}
References\hspace{\fill}9\\\vspace{1in}\maketitle

\section {Introduction}
It is well known that the quantum entanglement deteriorates very
fast due to environmental interactions and one would like to find
methods that can save or at least slow down the loss of entanglement
\cite{Nielsen}. It is also now known that quantum entanglement can
die much faster than the scale over which dephasing occurs
\cite{Yu}. For example the coherence of the qubit typically lasts
over a time scale of the order of $T_{2}$ whereas the entanglement
can exhibit sudden death and thus it is important to extend the
techniques used for single qubits to bipartite and even multipartite
systems. In this paper we examine how the pulse techniques which
were developed to examine the issue of dephasing can help in saving
the entanglement. The quantum dynamical decoupling
\cite{Viola,Ban,Facchi} uses a sequence of control pulses to be used
on the system at an interval much less than the time-scale of the
bath coherence time. In this way, the coupling of the system to the
bath can be time-reversed and thus canceled. Such non-Markovian
approach has been successfully applied  to two-level systems,
harmonic oscillators \cite{Vitali}. A different approach was used in
\cite{Agarwal1} where a control pulse was applied to a different
transition rather than the relevant two-level transition. This
technique shows that the control pulse causes destructive
interference between transition amplitudes at different times which
leads to inhibition of the spontaneous emission of an excited atom.
Similar techniques could be useful to suppress the decoherence of a
qubit coupled to a thermal bath. Other methods for protection
against dephasing are known. These include application of fast
modulations to the bath \cite{Agarwal2} as well as decoherence free
subspaces \cite{Palma}. The dynamical decoupling idea has been
implemented in a few recent experiments \cite{Kishimoto} with
excitons in semiconductors, with Rydberg atomic qubits, with solid
state qubits and with nuclear spin qubits.

More recent developments primarily due to Uhrig
\cite{Uhrig,Yang,GSUhrig,Lee} go far beyond than what has been done
earlier on dynamical decoupling. The dynamical decoupling schemes
use a series of $\pi$ pulses applied at regular interval of times.
The pulses reverse the evolution given by the Hamiltonian describing
the interaction with a dephasing environment. This is because under
a $\pi$ pulse the spin operator $S_{z}$ reverses sign. Uhrig
discovered that $\pi$ pulses applied at irregular intervals of time
are much more effective in controlling dephasing. The regular pulse
sequence and the Uhrig sequence are given by
\begin{equation}\label{1}
\displaystyle T_{j}=\frac{jT}{n+1},T_{j}=T\sin^{2}(\frac{\pi
j}{2(n+1)}).
\end{equation}

In this paper we focus on the utility of the sequence of pulses as
discovered by Uhrig in saving quantum entanglement. Unlike other
papers which focus on dephasing issues we concentrate on
entanglement. This is important as the dynamical behavior of
entanglement could be quite different than that of dephasing. We
calculate the concurrence parameter \cite{Wootters} which
characterizes the entanglement between the two qubits. We show the
net time evolution of the concurrence parameter under the action of
the Uhrig sequence of pulses and compare its evolution with the one
given by when the uniform sequence of pulses is applied. We show the
great advantage of the Uhrig sequence over the uniform sequence in
saving entanglement. A very recent experiment \cite{Du} establishes
the advantage of Uhrig's sequence in lengthening the dephasing time
of a single qubit. The organization of the paper is as follows: In
Sec 2 we introduce the microscopic model of dephasing and calculate
the relevant physical quantities under the influence of the control
pulses. In Sec 3 we show how the coherent state techniques can be
used to obtain the dynamical results. In Sec 4 we calculate the
dynamics of entanglement and present numerical results. In Sec 5 we
conclude with possible generalizations of our results on
entanglement.

\section{Dynamical decay of entanglement under dephasing}
Let us consider two qubits in an entangled state \cite{Yu} which in
general could be a mixed state. In terms of the basis states for the
two qubits, we choose the initial state as
\begin{equation}\label{2}
\begin{array}{lcl}
|1\rangle=|\uparrow\rangle_{A}\otimes|\uparrow\rangle_{B},|2\rangle=|\uparrow\rangle_{A}\otimes|\downarrow\rangle_{B},\vspace{.1in}\\
|3\rangle=|\downarrow\rangle_{A}\otimes|\uparrow\rangle_{B},|4\rangle=|\downarrow\rangle_{A}\otimes|\downarrow\rangle_{B}.
\end{array}
\end{equation}
\begin{equation}\label{3}
\rho=\left(
  \begin{array}{cccc}
    a & 0 & 0 & 0 \vspace{.05in}\\
    0 & b & z & 0 \vspace{.05in}\\
    0 & z^{*} & c & 0 \vspace{.05in}\\
    0  & 0 & 0 & d \vspace{.05in}\\
  \end{array}
\right).
\end{equation}
The state (\ref{3}) is positive and normalized if $a+b+c+d=1$ and
$bc>|z|^2$. This state has the structure of a Werner state. For
$a=d=0$; $b=c=|z|=1$, it represents a maximally entangled state. The
amount of entanglement in the state is given by the concurrence
given by
\begin{equation}\label{4}
\begin{array}{lcl}
C=\mbox{Max}\{0,\tilde{C}\},\vspace{0.1in}\\
\tilde{C}=2\{|\rho_{23}|-\sqrt{\rho_{11}\rho_{44}}\}=2|z|(1-r),\vspace{0.1in}\\\displaystyle
r=\frac{\sqrt{ad}}{|z|}.
\end{array}
\end{equation}
And therefore the state is entangled as long as $|z|$ is greater
than $\sqrt{ad}$. Now under dephasing the diagonal elements $a$ and
$d$ do not change. However the coherence in the qubit decays as
$\exp[-t/T_{2}]$ and therefore the entanglement survives as long as
$|z|\exp[-2t/T_{2}]-\sqrt{ad}>0$ and thus entanglement vanishes if
$\displaystyle t>\frac{T_{2}}{2}\ln{\frac{|z|}{\sqrt{ad}}}$.

We would now examine how the action of pulses can protect the
entanglement. We would calculate the time over which entanglement
can be made to survive. For this purpose we need to examine the
microscopic model of dephasing. We would make the reasonable
assumption that each qubit interacts with its own bath. We could
then examine the dynamics of the individual qubits and then obtain
the evolution of the concurrence.

On a microscopic scale the dephasing can be considered to arise from
the interaction of the qubit with a bath of oscillators i.e. from
the Hamiltonian
\begin{equation}\label{5}
\displaystyle H=\hbar\sum_{i}\omega_{i}a_{i}^{\dag}a_{i}+\hbar
S_{z}\sum_{i}g_{i}(a_{i}+a_{i}^{\dag}),
\end{equation}
where the $S_{z}$ is the $z$ component of the spin operator for the
qubit and the annihilation and creation operators $a_{i}$,
$a_{i}^{\dag}$ represent the oscillators of the Bosonic bath. The
bath is taken to have a broad spectrum. In particular for an Ohmic
bath we take spectrum of bath as
\begin{equation}\label{6}
\displaystyle
J\rightarrow\sum_{i}|g_{i}|^2\delta(\omega-\omega_{i})=2\alpha\omega\Theta(\omega_{D}-\omega),
\end{equation}
where $\omega_{D}$ is the cut off frequency. It essentially
determines the correlation time of the bath. Such a bath leads to
dephasing i.e. the spin polarization decays at the rate $T_{2}$. The
dynamical decoupling schemes use a series of $\pi$ pulses applied at
regular interval of times whereas Uhrig applies $\pi$ pulses at
irregular intervals of time. Such nonuniformly spaced pulses are
much more effective in controlling dephasing over a time interval determined by the cut off frequency and number of pulses. The regular sequence of pulses is more effective outside this domain. The pulses reverse the
evolution given by the interaction part in the Hamiltonian (\ref{5})
since under a $\pi$ pulse the spin operator $S_{z}$ reverses sign.
The regular pulse sequence and the Uhrig sequence are given by
equation (\ref{1}). We need to calculate the dynamical evolution of
the off diagonal element of the density matrix for the qubit. We
work in the interaction picture hence the Hamiltonian (\ref{5})
becomes
\begin{equation}\label{7}
\displaystyle H=\hbar
S_{z}\sum_{i}g_{i}(a_{i}e^{-i\omega_{i}t}+a_{i}^{\dag}e^{i\omega_{i}t})=\hbar
S_{z}B(t),
\end{equation}
where $B(t)$ is the bath operator given by
\begin{equation}\label{8}
\displaystyle
B(t)=\sum_{i}g_{i}(a_{i}e^{-i\omega_{i}t}+a_{i}^{\dag}e^{i\omega_{i}t}),
\end{equation}

It is easy to see that the off diagonal element of the single qubit
density matrix $\sigma$ is
\begin{equation}\label{9}
\displaystyle \sigma_{\uparrow\downarrow}(t)=\Tr_{B}\langle
\downarrow|U(t)\sigma_{B}\sigma(0)U^{\dag}(t)|\uparrow\rangle,
\end{equation}
where $\Tr_{B}$ is over the initial bath density matrix $\sigma_{B}$
and where
\begin{equation}\label{10}
\displaystyle U(t)=T\exp\{-i\int_{0}^{t}S_{z}B(\tau)d\tau\}.
\end{equation}
This can be simplified to
\begin{equation}\label{11}
\begin{array}{lcl}
\sigma_{\uparrow\downarrow}(t)=\sigma_{\uparrow\downarrow}(0)
\Tr_{B}
V_{-}(t)\sigma_{B}V_{+}^{\dag}(t),\vspace{0.1in}\\\hspace{0.45in}=\sigma_{\uparrow\downarrow}(0)\langle
V_{+}^{\dag}(t)V_{-}(t)\rangle,
\end{array}
\end{equation}
where
\begin{equation}\label{12}
\displaystyle V_{\pm}(t)=T\exp\{\mp
\frac{i}{2}\int_{0}^{t}B(\tau)d\tau\}.
\end{equation}
Thus we can write
\begin{equation}\label{13}
\sigma_{\uparrow\downarrow}(t)=\sigma_{\uparrow\downarrow}(0)\zeta(t),
\end{equation}
\begin{equation}\label{14}
\begin{array}{lcl}
\zeta(t)=\langle
V_{+}^{\dag}(t)V_{-}(t)\rangle,\vspace{0.1in}\\\hspace{0.3in}=\langle
T\exp\{i\int_{0}^{t}B(\tau)d\tau\}\rangle.
\end{array}
\end{equation}
\begin{figure}
\begin{center}
\scalebox{0.8}{\includegraphics{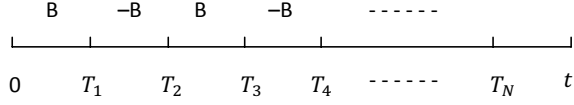}} \caption{\label{Fig1} A
sequence of $\pi$ pulses is applied at times $T_{j}$.}
\end{center}
\end{figure}
So far no approximation has been made. Now we incorporate the effect
of pulses in the dynamical evolution of the single qubit coherence.
Let us apply a sequence of $\pi$ pulses at times $T_{j}$ as shown in
figure 1. At each $T_{j}$ the interaction Hamiltonian changes sign.
This can be easily incorporated in the dynamics and the result is
\begin{equation}\label{15}
\begin{array}{lcl}
\zeta(t)=\langle W(t)\rangle,\vspace{0.1in}\\
W(t)=T\exp\{i\int_{0}^{t}B(\tau)f(\tau)d\tau\},\vspace{0.1in}\\
\displaystyle
f(t)=\sum_{j=0}^{N-1}(-1)^{j}\theta(t-T_{j})\theta(T_{j+1}-t),
\end{array}
\end{equation}
where the step function $\theta(t)=1$ if $t>0,=0$, if $t<0$. It is
especially instructive to use coherent state techniques to simplify
the expression for $W$. We do this in the next section.
\section{Single Qubit Coherence $S(t)$}
We now examine the calculation of the function $W(t)$. We note that
the bath operator $B(\tau)$ is such that the commutator
$[B(\tau_{1}),B(\tau_{2})]$ is a c-number. In such a case it has
been shown by Glauber \cite{Glauber1} that the time ordering can be
simplified. It can be shown that
\begin{equation}\label{16}
\fl W=\exp\{i\int_{0}^{t}B(\tau)d\tau
f(\tau)\}\exp\{-\frac{1}{2}\int_{0}^{t}d\tau_{1}\int_{0}^{\tau_{1}}d\tau_{2}[B(\tau_{1}),B(\tau_{2})]f(\tau_{1})f(\tau_{2})\}.
\end{equation}
Since $B$ is a Hermitian operator, the last exponential is just a
c-number phase factor and hence
\begin{equation}\label{17}
W=\exp(i\Phi(t))\exp\{i\int_{0}^{t}B(\tau)d\tau f(\tau)\},
\end{equation}
\begin{equation}\label{18}
\hspace{0.2in}=\exp(i\Phi(t))\Pi_{j}\exp\{if_{j}a_{j}+if_{j}^{*}a_{j}^{\dag}\},
\end{equation}
where
\begin{equation}\label{19}
f_{j}=g_{j}\int_{0}^{t}e^{-i\omega_{j}\tau}f(\tau)d\tau.
\end{equation}
On using the Baker-Hausdorff identity (\ref{18}) can be further
simplified to
\begin{equation}\label{20}
W=\exp(i\Phi(t))\Pi_{j}\exp\{if_{j}^{*}a_{j}^{\dag}\}\exp\{if_{j}a_{j}\}\exp\{-\frac{1}{2}|f_{j}|^2\},
\end{equation}
The thermal expectation value of $W$ can be easily obtained using
for example the P-representation for the thermal density matrix
\cite{Glauber2}
\begin{equation}\label{21}
\begin{array}{lcl}
\displaystyle \rho_{th j}=\frac{1}{\pi n_{j}}\int
\exp\{-\frac{|\alpha|^2}{n_{j}}\}|\alpha\rangle\langle\alpha|d^{2}\alpha,\vspace{0.1in}\\\hspace{0.2in}
\displaystyle n_{j}=\frac{1}{e^{\beta\hbar\omega_{j}}-1}.
\end{array}
\end{equation}
Thus
\begin{equation}\label{22}
\fl
W=\exp(i\Phi(t))\Pi_{j}\exp\{-\frac{1}{2}|f_{j}|^2\}\displaystyle\times\frac{1}{\pi
n_{j}}\int
\exp\{-\frac{|\alpha|^2}{n_{j}}\}\exp\{if_{j}^{*}\alpha^{*}+if_{j}\alpha\}d^{2}\alpha,\vspace{0.1in}\\
\end{equation}
which on simplification reduces to
\begin{equation}\label{23}
W=\exp(i\Phi(t))\Pi_{j}\exp\{-(n_{j}+\frac{1}{2})|f_{j}|^2\},
\end{equation}
On using the form of $f_{j}$ and on introducing the spectral density
of the bath oscillators, the expression (\ref{23}) becomes
\begin{equation}\label{24}
W=\exp(i\Phi(t))\exp\{-\int
d(\omega)J(\omega)[n(\omega)+\frac{1}{2}]|f(\omega)|^{2}\},
\end{equation}
where now
\begin{equation}\label{25}
f(\omega)=\int_{0}^{t}e^{-i\omega\tau}F(\tau)d\tau.
\end{equation}
The function $f$ can be simplified using the explicit form of
$F(\tau)$:
\begin{equation}\label{26}
\displaystyle f(\omega)=-i[1+(-1)^{N+1}e^{-i\omega
t}+2\sum_{j=1}^{N}(-1)^{j}e^{-i\omega T_{j}}].
\end{equation}
The result (\ref{23}) is equivalent to equation (\ref{8}) of Uhrig. We also note that results like (24) appear in the earlier literature \cite{Agarwal2} dealing especially with nonmarkovian master equations.
\section{Saving Entanglement: Numerical Results}
Since we work under the assumption that each qubit interacts with
its own bath, the time dependent matrix elements of the density
matrix in the basis (\ref{2}) can be obtained by noting that the
diagonal elements do not evolve under dephasing. The off diagonal
element $\rho_{23}(t)$ is given by
\begin{equation}\label{27}
\rho_{23}(t)=\rho_{23}(0)|\zeta(t)|^2,
\end{equation}
where $\zeta(t)$ is defined by equation (\ref{13}) and its explicit
form is given by equation (\ref{23}). Thus
\begin{equation}\label{28}
|\zeta(t)|^2=S(t)=\exp\{-2\int d\omega J(\omega)
(n(\omega)+\frac{1}{2})|f(\omega)|^{2}\},
\end{equation}
It can then be shown that the time dependence of the concurrence is
given by
\begin{equation}\label{29}
\begin{array}{lcl}
C(t)=\mbox{Max}\{0,\tilde{C}(t)\},\vspace{0.1in}\\
\displaystyle \tilde{C}(t)=2|z|\{S(t)-r\},\vspace{0.1in}\\
\displaystyle r=\frac{\sqrt{ad}}{|z|}.
\end{array}
\end{equation}

\begin{figure}
\begin{center}
\scalebox{0.8}{\includegraphics{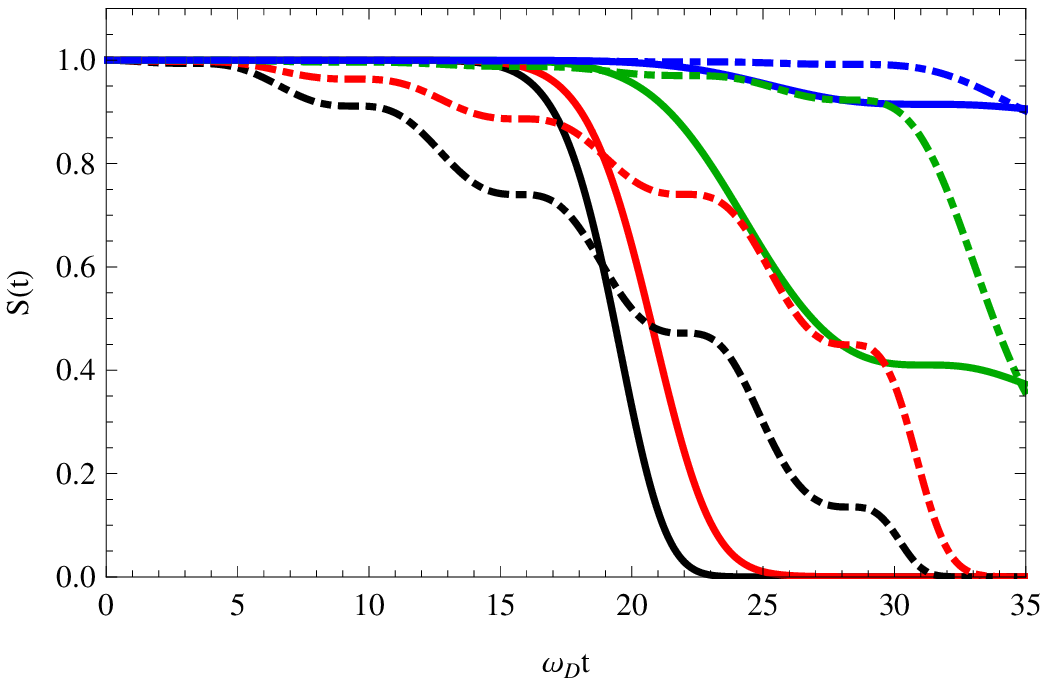}} \caption{\label{Fig2} Signal
vs time for $n=10$ at $T=0$. Solid lines for the optimized sequence,
dotdashed lines for the equidistant sequence. From bottom to top the
curves correspond to $\alpha=0.25,0.1,0.01,0.001$.}
\end{center}
\end{figure}
\begin{figure}
\begin{center}
\scalebox{0.8}{\includegraphics{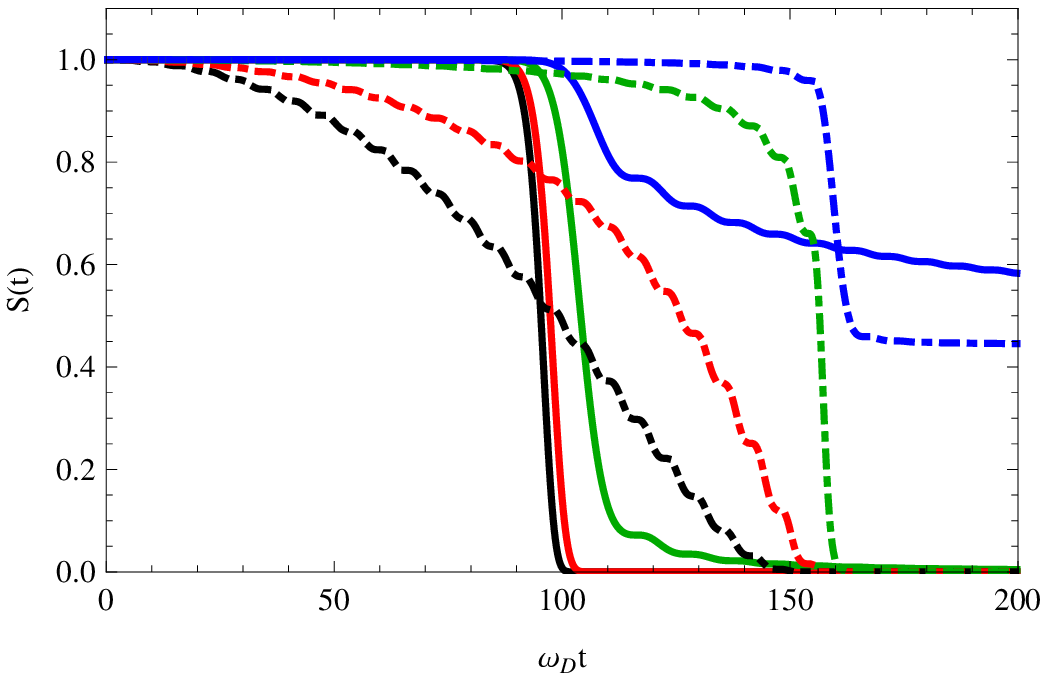}} \caption{\label{Fig3} Signal
vs time for $n=50$ at $T=0$. Solid lines for the optimized sequence,
dotdashed lines for the equidistant sequence. From bottom to top the
curves correspond to $\alpha=0.25,0.1,0.01,0.001$.}
\end{center}
\end{figure}
We next discuss the dynamical behavior of the entanglement. The
function $f(\omega)$ has been evaluated by Uhrig.  For the pulses
applied at regular intervals and for $n$ even we have
\begin{equation}\label{30}
|f(\omega)|^2=4\tan^2[\omega t/(2n+2)]\cos^{2}(\omega
t/2)/\omega^{2}\hspace{0.1in} \forall\hspace{0.1in} n\hspace{0.1in}
\rm{even},
\end{equation}
whereas for the Uhrig's pulse sequence
\begin{equation}\label{31}
|f(\omega)|^{2}\approx 16(n+1)^{2}J_{n+1}^{2}(\omega/2)/\omega^{2},
\end{equation}
where $J_n$ is the Bessel function. The function $S(t)$ is shown in
figures 2 and 3 for $n=10$ and $n=50$. The parameter
$\omega_{D}^{-1}$ is a measure of the bath correlation time. These
figures show that the entanglement lives much longer for Uhrig
sequence of pulses applied at nonuniform intervals of time provided that $\omega_{D}t\leq2n$. Thus entanglement can be made
to live over times which could be several orders longer than the
coherence time of the bath.

\section{Conclusions}

In conclusion we have considered how the effects of dephasing on the
destruction of entanglement  can be considerably slowed on by
applying the sequence of $\pi$ pulses applied at time intervals
given by Uhrig. We demonstrated this explicitly for the case of a
mixed entangled state of two qubits. The sequence given by Uhrig is
far better in controlling the death of entanglement compared to the
sequence applied at regular intervals of time. These conclusions
also apply to the multiparticle entangled state like the GHZ state
\begin{equation}\label{32}
\displaystyle
|\Psi\rangle=\frac{1}{\sqrt{2}}(|\uparrow\cdots\uparrow\rangle-|\downarrow\cdots\downarrow\rangle),
\end{equation}
whose entanglement under dephasing would decay as the density matrix
at time $t$ would be
\begin{equation}\label{33}
\begin{array}{lcl}
\displaystyle
\rho(t)=\frac{1}{2}|\uparrow\cdots\uparrow\rangle\langle\uparrow\cdots\uparrow|+\frac{1}{2}|\downarrow\cdots\downarrow\rangle\langle\downarrow\cdots\downarrow|
\vspace{0.1in}\\\hspace{0.45in}\displaystyle
-\frac{1}{2}\exp\{-\frac{tN}{T_{2}}\}(|\uparrow\cdots\uparrow\rangle\langle\downarrow\cdots\downarrow|+c.c.).
\end{array}
\end{equation}
Under the application of $\pi$ pulses, the prefactor $\displaystyle
\exp\{-\frac{tN}{T_{2}}\}$ would be replaced by $(S(t))^{N/2}$.
Since $S(t)$ can be made close to unity for times even orders of the
correlation time of the bath, the entanglement of the multiparticle
GHZ state would survive over a long time. Note further that Uhrig's
work has been generalized to arbitrary relaxations \cite{Yang}.
Clearly these generalizations should be applicable to the
considerations of entanglement. In particular we hope to examine the
protection of Werner state against different models of environment.

Finally we note that our ongoing work also suggests how other
methods like photonic crystal environment can be used to save
entanglement.

\end{document}